\documentclass[showpacs,preprintnumbers,amsmath,amssymb]{revtex4}
\usepackage{graphicx}
\usepackage{dcolumn}
\usepackage{bm}

\begin{document}
\draft

\title{Beatwave Excitation of Plasma Waves Based on Relativistic Bi-Stability}

\author{Gennady Shvets}
\affiliation{Department of Physics and Institute for Fusion
Studies, The University of Texas at Austin, Austin, TX 78712}

\newcommand{\ba}{\begin{eqnarray}}
\newcommand{\ea}{\end{eqnarray}}
\newcommand{\be}{\begin{equation}}
\newcommand{\ee}{\end{equation}}
\newcommand{\para}{\parallel}

\begin{abstract}
A nonlinear beatwave regime of plasma wave excitation is
considered. Two beatwave drivers are considered:
intensity-modulated laser pulse and density-modulated
(microbunched) electron beam. It is shown that a long beatwave
pulse can excite strong plasma waves in its wake even when the
beatwave frequency is detuned from the electron plasma frequency.
The wake is caused by the dynamic bi-stability of the nonlinear
plasma wave if the beatwave amplitude exceeds the analytically
calculated threshold. In the context of a microbunched beam driven
plasma wakefield accelerator, this excitation regime can be
applied to developing a femtosecond electron injector.
\end{abstract}


\maketitle

Beatwave excitation of electron plasma waves continues attracting
significant attention as a basic nonlinear plasma phenomenon, and
as a viable approach to plasma-based particle
acceleration~\cite{tajima_dawson,joshi_kats,tochitsky_prl04,wurtele_04}.
Beatwave excitation mechanism is realized when the driver
intensity (laser or particle beam) is modulated with the temporal
periodicity of the plasma wave. The linear one-dimensional theory
of the beatwave-driven plasma wave generation is well
understood~\cite{tajima_dawson,esarey_review96}, and its most
important predictions are as follows. First, the effectiveness of
plasma wave excitation is strongly dependent on the difference
between the beatwave frequency $\omega_B$ and plasma wave
frequency $\omega_p = \sqrt{4\pi e^2 n_0/m}$ (where $-e$ and $m$
are the electron charge and mass, and $n_0$ is the plasma
density): the smaller is the frequency detuning $\Delta \omega
\equiv \omega_B - \omega_p$, the larger is the resulting plasma
wave inside the beatwave. Second, only if the beatwave pulse is
short enough for its bandwidth to be comparable to $\Delta
\omega$, an appreciable plasma wave is left in its wake.

In this Letter I demonstrate that these conclusions are no longer
valid when the relativistic nonlinearity of a plasma wave is
accounted for. In particular, a strong plasma wave can be excited
in the wake of a relatively long beatwave pulse of duration $t_L
\gg 1/\Delta \omega$ due to the nonlinear phenomenon of dynamic
relativistic bi-stability (RB)~\cite{kaplan82}. Another
manifestation of RB is that, at a certain critical strength of the
beatwave driver, a weak driven plasma wave becomes unstable, and a
much higher amplitude wave is excited. Linear estimates of the
plasma wave amplitude fail when the beatwave amplitude exceeds
this detuning-dependent critical strength. As the time-dependent
beatwave strength increases and exceeds the critical value,
significant pulsations of the plasma wave amplitude occur. These
pulsations indicate that significant energy exchange takes place
between the plasma wave and the driver. This effect can be
exploited in a plasma wakefield accelerator driven by a
microbunched electron beam~\cite{liu_prl98}: bunches in the head
of the beam excite while those in the back deplete plasma waves,
thereby gaining energy.

Relativistic bi-stability was originally described~\cite{kaplan82}
for a magnetized electron subjected to cyclotron heating.
Applications of RB to electron cyclotron heating of fusion
plasmas~\cite{nevins87,stupakov90} have been later suggested.
Although the nonlinear nature of electron plasma waves has been
noted
before~\cite{rosenbluth_liu,tang_sprangle85,mckinstrie87,meerson91},
the RB of plasma waves has not been explored, either as a basic
phenomenon or in the context of plasma-based accelerators.

The one-dimensional relativistic dynamics of the cold plasma
driven by a beatwave can be described using a Lagrangian
displacement of the plasma element originally located at $z_0$:
$z(t) = z_0 + \zeta(t,z_0)$. It is assumed that the beatwave
generated by either a pair of frequency-detuned laser beams, or a
modulated electron beam, is moving with the speed close to the
speed of light $c$, and, therefore, all beatwave quantities are
functions of the co-moving coordinate $\tau^{\prime} = \omega_p(t
- z/c) \equiv \tau - \omega_p \zeta/c$. Introducing the normalized
displacement $\tilde{\zeta} = \omega_p \zeta/c$ and longitudinal
relativistic momentum $\tilde{p} = \gamma d \tilde{\zeta}/d \tau$,
where $\gamma = \sqrt{1 - \vec{v}^2/c^2}$, equations of motion
take on the form
\begin{equation}\label{eq:motion1}
    \frac{d \tilde{\zeta}}{d \tau} =
    \frac{\tilde{p}}{\sqrt{1 + \tilde{p}^2}}, \ \ \ \ \
    \frac{d \tilde{p}}{d \tau} = - \tilde{\zeta} + a(\tau^{\prime}) \cos{\omega
    \tau^{\prime}}.
\end{equation}
Assuming that $|\Delta \omega| \ll \omega_p$ (near-resonance
excitation), transverse momentum of the plasma has been neglected
and the relativistic $\gamma$-factor simplified to $\gamma =
\sqrt{1+\tilde{p}^2}$. The first term in the force equation is the
restoring force of the ions, and the second term signifies the
beatwave with the frequency $\omega_B \equiv \omega \omega_p$. The
nonlinear in $\zeta$ modification of the beatwave in the rhs of
Eqs.~(\ref{eq:motion1}) is neglected in what follows. For a pair
of linearly polarized laser pulses with electric field amplitudes
$E_1$ and $E_2$ and the corresponding frequencies $\omega_1$ and
$\omega_2 = \omega_1 - \omega_B$ the normalized beatwave amplitude
$a = (e/mc)^2 E_1 E_2/2\omega_1 \omega_2$~\cite{rosenbluth_liu}.
For a driving electron bunch with the density profile $n_b =
n_{b0} + \delta n_b \sin{\omega \tau}$ it can be shown that $a =
\delta n_b/n_0$. Although arbitrary profiles of $a(\tau)$ are
allowed, it is assumed that $|da/d\tau | \ll |a|$. The total
energy density of the plasma wave $U_p/n_0 mc^2 =
\sqrt{1+\tilde{p}^2} + \tilde{\zeta}^2/2$ is changed via the
interaction with the beatwave. The effect of the plasma wave on
the beatwave is neglected for the moment and addressed towards the
end of the Letter.

Although Eqs.~(\ref{eq:motion1}) can be solved numerically at this
point, further simplification is made by assuming $\tilde{p} = u
\cos{(\omega \tau + \phi)}$, where $u$ and $\phi$ are slowly
varying functions of $\tau$. In the weakly relativistic
approximation $\tilde{p}^2 \ll 1$ obtain:
\begin{eqnarray}
    &&\frac{d u}{d \tau} = \frac{a}{2} \cos{\phi} \label{eq:dudt}
    \\ &&u \frac{d \phi}{d \tau} = -\frac{a}{2} \sin{\phi} -
    \frac{u}{2\omega}(\omega^2 - 1 + 3u^2/8). \label{eq:dphidt}
\end{eqnarray}

\begin{figure}
\includegraphics[height=7cm]{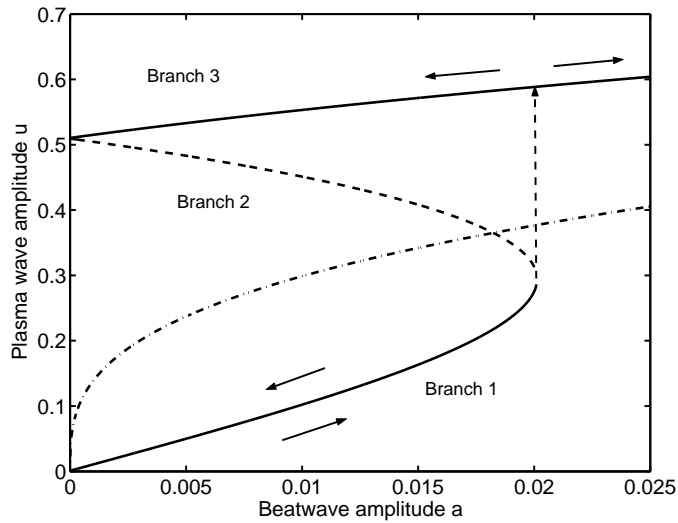}
\caption{\label{fig:steady_hyster} Steady-state solutions of a
driven plasma wave as a function of the beatwave amplitude $a$.
Solid lines $1$, $2$: stable equilibria for $\omega = 0.95$;
dashed line: unstable equilibrium for $\omega = 0.95$; dot-dashed
line: resonant excitation with $\omega = 1$.}
\end{figure}

Depending on the beatwave frequency $\omega$ and the amplitude
$a$, equilibrium solutions $du/d\tau = 0$ (steady amplitude) and
$d\phi/d\tau = 0$ (phase-locking to the beatwave) of
Eqs.~(\ref{eq:dudt},\ref{eq:dphidt}) can have one or three real
roots. For any $\omega$ there is a stable equilibrium point:
$\phi_0 = -\pi/2$ and $u_0 > 0$ found as the root of the
third-order polynomial equation $\cal{P}$$(u_0) = u_0 (\omega^2 -1
+ 3/8 u_0^2) = \omega a$. For the most interesting $\omega < 1$
regime additional solutions $\phi_0 = \pi/2$ and $u_0 > 0$, where
$u_0$ is the positive root of $\cal{P}$ $(u_0) = - \omega a$, may
be found, depending on the beatwave amplitude. Specifically, there
are no additional positive roots for $a > a_{crit}$, where
$a_{crit} = 4\sqrt{2} (1-\omega^2)^{3/2}/9\omega$, and two
positive roots $u_{1,2}$ for $a < a_{crit}$ (one of them
unstable). Stable equilibrium amplitudes $u_0$ with $\phi_0 =
\pi/2$ (Branch 1) and $\phi_0 = -\pi/2$ (Branch 3), as well as the
unstable one (Branch 2) are plotted in
Fig.~\ref{fig:steady_hyster} as a function of the beatwave
strength $a$ for $\omega = 0.95$ ($a_{crit} = 0.02$). Equilibrium
bi-stability corresponding to Branches 1 and 3 is universal for
any nonlinear pendulum~\cite{landau_mech,kaplan82}, including a
weakly damped one. Equilibrium solutions are meaningful only if
the plasma wave is phase-locked to the beatwave: $d\phi/d\tau
\approx 0$. As shown below, this is not the case when the peak
beatwave amplitude exceeds $a_{crit}$. Nonetheless, a {\it
dynamic} RB described below occurs even in the absence of
phase-locking.


\begin{figure}
\includegraphics[height=7cm]{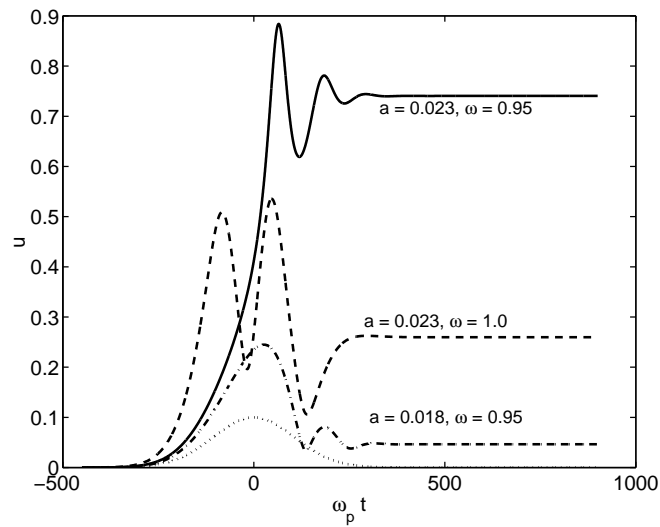}
\caption{\label{fig:beatwave_fig2} Excitation of a plasma wave by
a Gaussian beatwave pulse (dotted line), $a(\tau) = a_0
\exp{[-\tau^2/\tau_L^2]}$, $\tau_L = 150$. Solid line: $\omega =
0.95$, above-threshold excitation with $a_0 = 0.023 > a_{crit} =
0.02$; dashed line: resonant excitation with $\omega = 1$ and $a_0
= 0.023$; dot-dashed line: $\omega = 0.95$, below-threshold
excitation with $a_0 = 0.018 < a_{crit}$.}
\end{figure}

Consider plasma response to a Gaussian beatwave pulse $a(\tau) =
a_0 \exp{(-\tau^2/\tau_L^2)}$, where $\tau_L \gg 1/|1 - \omega|$
is the normalized pulse duration. For $a_0 < a_{crit}$ the plasma
response is as follows: amplitude $u$ adiabatically follows
$a(\tau)$ by staying on the Branch $1$ and following the
equilibrium trajectory schematically shown by arrows in
Fig.~\ref{fig:steady_hyster}. The adiabaticity condition is
$\Omega_B \tau_L \gg 1$, where $\Omega_B$ is the bounce frequency
around the equilibrium point $u_0$ such that $\cal{P}$$(u_0) =
-\omega a(\tau)$. Linearizing Eqs.~(\ref{eq:dudt},\ref{eq:dphidt})
around $\phi = \pi/2$ and $u = u_0$ yields $\Omega_B^2 =a
(u_{crit}^2 - u_0^2)/4\omega u_0$, where $u_{crit} = 2\sqrt{2(1 -
\omega^2)}/3$ is the critical plasma wave amplitude corresponding
to the merging point between Branches $1$ and $2$ in
Fig.~\ref{fig:steady_hyster}. For $a_0 < a_{crit}$ plasma
oscillation is indeed phase-locked to the beatwave at $\phi_0
\approx \pi/2$ during the ramp-up and most of the ramp-down of the
laser pulse (although phase-locking is lost when the pulse
amplitude becomes very small on the down-ramp). As the result,
plasma wave amplitude returns to a very small value in the wake of
the beatwave, as shown by a dot-dashed line in
Fig.~\ref{fig:beatwave_fig2}. The longer is the beatwave pulse
duration $\tau_L$, the smaller is the wake because its
non-vanishing amplitude is due to the adiabaticity violation for
finite $\tau_L$.

Situation changes for $a_0 > a_{crit}$: as $a(\tau)$ approaches
$a_{crit}$, the adiabatic condition is violated (noted in the
context of electron cyclotron heating~\cite{nevins87,stupakov90}),
and phase-locking at $\phi_0 = \pi/2$ is no longer possible. Thus,
the transfer to Branch $3$ schematically shown by a vertical arrow
in Fig.~\ref{fig:steady_hyster} becomes feasible, and the plasma
wave amplitude can dramatically increase. In the presence of a
finite plasma wave damping this indeed happens: the subsequent
decrease of the beatwave amplitude results in phase-locking at
$\phi_0 = -\pi/2$, with $u$ following along the Branch $3$.
Without damping, there is no mechanism for the plasma wave to
reach the equilibrium amplitude given by the upper Branch $3$. As
shown below, a conservation law prohibits the jump between
Branches $1$ and $3$.

Nevertheless, even without damping, a significant plasma wave is
left behind the finite-duration beatwave pulse
(Fig.~\ref{fig:beatwave_fig2}, solid line). The previously
unaccessible finite-amplitude solution has been reached due to the
effect of the dynamic RB which is best understood through the
conservation of the effective Hamiltonian of the driven plasma
wave. The effective Hamiltonian
\begin{equation}\label{eq:hamilt}
    H = \frac{1}{2} a u \sin{\phi} + \frac{(\omega^2 -
    1)u^2}{4\omega} + \frac{3u^4}{64 \omega}
\end{equation}
can be used to express Eqs.~(\ref{eq:dudt},\ref{eq:dphidt}) in the
form of $\dot{u} = (1/u) dH/d\phi$, $\dot{\phi} = -(1/u) dH/du$.
For a slowly changing beatwave amplitude $a(\tau)$ the Hamiltonian
is almost conserved: $dH/d\tau = 0.5 u \sin{\phi} da/d\tau \approx
0$. This constitutes the conservation law preventing the jump
between Branches $1$ and $3$. For the initially quiescent plasma
$a = 0$ and $u = 0$ before the arrival of the beatwave. Therefore,
$H \approx 0$ after its passage, as confirmed by numerical
simulations of various pulse durations and amplitudes. Remarkably,
in addition to the trivial quiescent plasma solution $u = 0$,
there is a second $u_{\infty} = 4\sqrt{(1-\omega^2)/3}$ solution
satisfying $H(u_{\infty}) = 0$. Thus, a plasma wave with $H = 0$
is dynamically bi-stable: after the passage of the beatwave it can
be either quiescent, or have the finite amplitude $u_{\infty}$. It
is conjectured that, by using a beatwave pulse with $a_0 >
a_{crit}$, the latter solution can be accessed, thereby leaving a
wake of a substantial plasma wave with amplitude $u_{\infty}$.

\begin{figure}
\includegraphics[height=7cm]{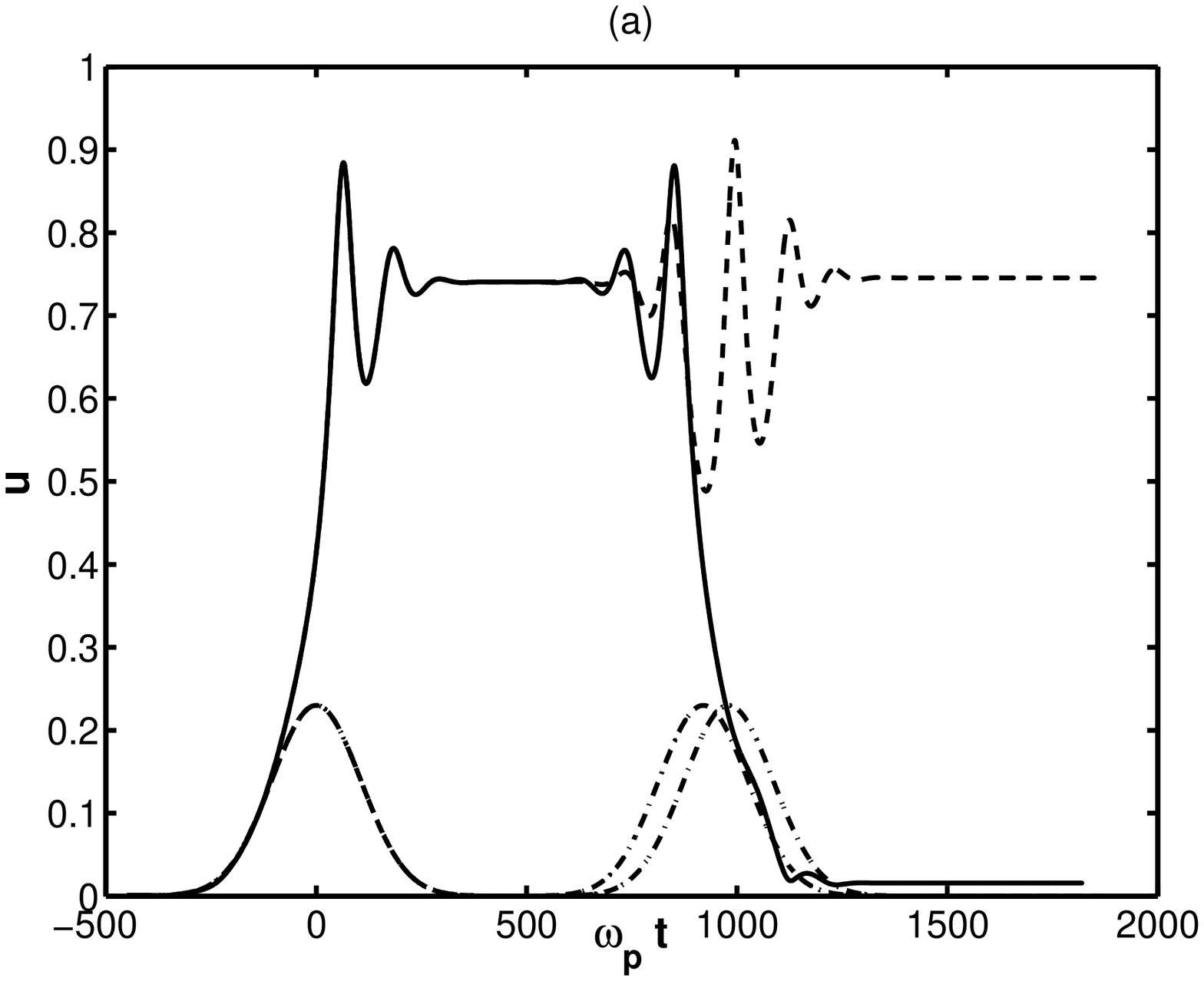}
\includegraphics[height=7cm]{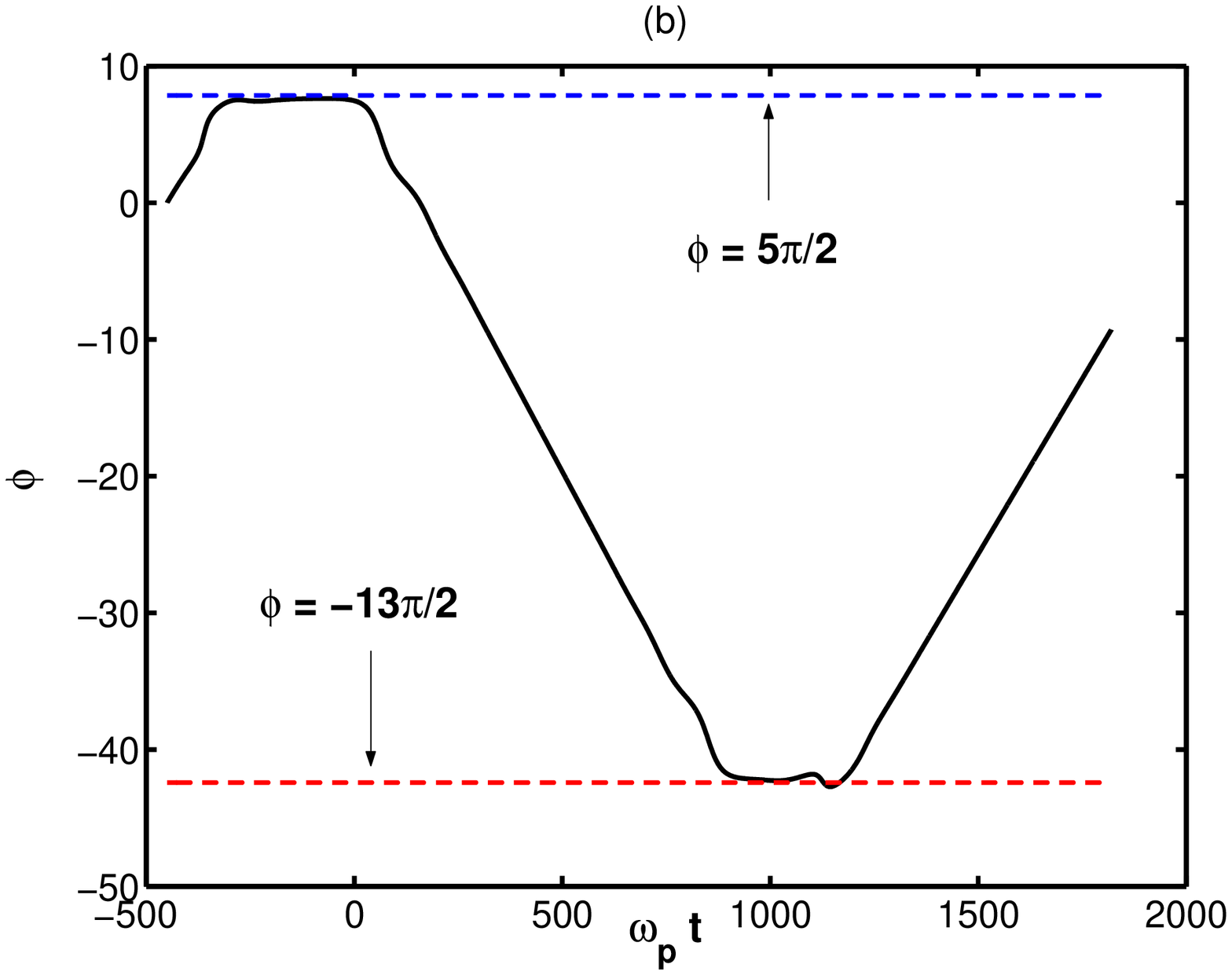}
\caption{\label{fig:beatwave_fig3} (a) Excitation of a plasma wave
by a pair of identical Gaussian beatwave pulses (dot-dashed line)
separated by the delay times $\tau_d = 920$ (solid line: wake
depleted by the second pulse) and $\tau_d = 980$ (dashed line:
wake unperturbed by the second pulse). Pulse parameters: same as
in Fig.~\ref{fig:beatwave_fig2}: $a_0 = 0.023$, $\omega = 0.95$,
and $\tau_L = 150$. (b) Sequence of phase lockings and phase
releases for $\tau_d = 920$.}
\end{figure}

This conjecture is verified by numerically integrating
Eqs.~(\ref{eq:dudt},\ref{eq:dphidt}) for two different detunings
(resonant, with $\omega = 1$, and non-resonant, with $\omega =
0.95$) and two beatwave amplitudes (sub-threshold, with $a_0 =
0.018$, and above-threshold, with $a_0 = 0.023$). In all cases the
Gaussian pulse duration was chosen $\tau_L = 150$. In physical
units, for the plasma density of $n_0 = 10^{19}$cm$^{-3}$ the
corresponding pulse duration is $t_L \equiv \tau_L/\omega_p
\approx 750$ fs. Simulation results are shown in
Fig.~\ref{fig:beatwave_fig2}, where the solid line corresponds to
the most interesting of the three cases: $\omega = 0.95$ and $a_0
= 0.023$. The plasma wave amplitude of $u \approx 0.75$ in the
wake of the laser pulse is in a good agreement with $u_{\infty} =
0.72$. This wake owes its existence to the dynamic RB: upon
interacting with the above-threshold laser beatwave, plasma wave
is transferred  from the quiescent state of $u = 0$ to the excited
state of $u = u_{\infty}$. The sub-threshold excitation
(dot-dashed line) with the same detuning fails to transfer the
plasma into the excited state, yielding a negligible wake that is
an order of magnitude smaller than in the above-threshold regime.

\begin{figure}
\includegraphics[height=7cm]{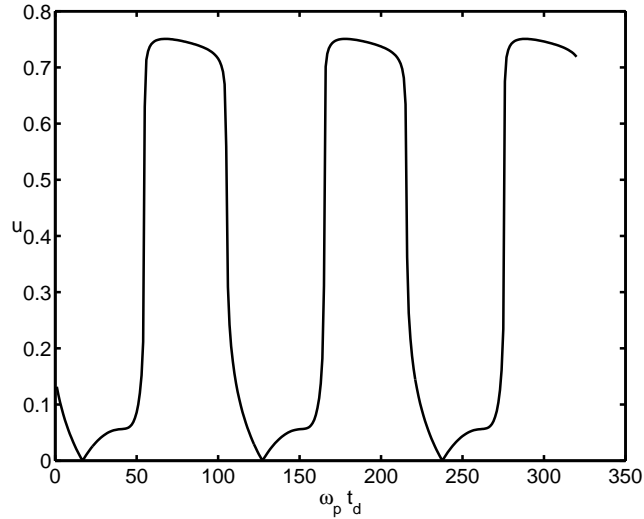}
\caption{\label{fig:beatwave_fig4} Dependence of the residual
plasma wave amplitude $u$ in the wake of a pair of identical
beatwave pulses on the time delay between the pulses $t_d$. Zero
corresponds to $t_d = 900/\omega_p$. Pulse parameters are the same
as in Fig.~\ref{fig:beatwave_fig3}.}
\end{figure}

Linear theory also fails to describe the strong wake in this
example because the detuning and the pulse duration are chosen
such that the linear prediction $u_{lin} = a_0/(1 -
\omega^2)\times \exp{[-\tau_L^2 (\omega - 1)^2/4]} \approx 0$ is
negligibly small. Resonant excitation (dashed line) also yields a
much smaller wave. Moreover, the resonantly and the sub-threshold
excited plasma waves would have been even smaller had the
adiabatic assumption been fully satisfied. Indeed, it is
numerically confirmed that the wake amplitudes for the resonant
and the sub-threshold excitations rapidly decline for longer
pulses, whereas the amplitude of the non-resonant above-threshold
excitation is insensitive to the beatwave pulse length $\tau_L$.

To demonstrate the bi-stable nature of the relativistic plasma
wake, excitation by a {\it pair} of identical beatwave pulses is
considered. By varying the time delay $\tau_d \equiv \omega_p t_d$
between the pulses, plasma wave can be either returned to the
original quiescent state $u = 0$ (Fig.~\ref{fig:beatwave_fig3}(a),
solid line, delay time $\tau_d = 920$), or brought into the
excited state $u = u_{\infty}$ (Fig.~\ref{fig:beatwave_fig3}(a),
dashed line, delay time $\tau_d = 980$). Depending on the time
delay $\tau_d$, there are, essentially, only two outcomes for
plasma wave amplitude: $u \approx 0$ or $u = u_{\infty}$. This
result is remarkably nonlinear: the linear theory predicts that
the wake behind two pulses depends on their separation in a
sinusoidal way: $u(t=\infty) = 2 u(t_1) \cos^2{[\pi \tau_d (\omega
- 1)]}$, where $\tau_L \ll t_1 \ll \tau_d$ is the instance well
after the end of the first and before the beginning of the second
pulse. The dependence of $u(t=\infty)$ on the delay time plotted
in Fig.~\ref{fig:beatwave_fig4} illustrates the effect for the
identical Gaussian pulses with $\tau_L = 150$, $a_0 = 0.023$, and
$\omega = 0.95$.

Dynamical RB described in this Letter is different from the
standard equilibrium bi-stablity of a weakly-damped nonlinear
oscillator~\cite{kaplan82,landau_mech} in that the former does not
require phase-locking, only the conservation of the effective
Hamiltonian $H$. As Fig~\ref{fig:beatwave_fig3}(b) indicates,
phase locking at $\phi_0 = \pi/2$ exists only during the switch-on
half of the beatwave, $-2 \tau_L < \tau < 0$. The plasma wave
phase is released afterwards ($0 < \tau < \tau_d$) as the pulse
amplitude settles into $u = u_{\infty}$. Depending on the delay
time, the second pulse can either (i) lock the phase at $\phi_0 =
\pi/2$ (as shown in Fig~\ref{fig:beatwave_fig3}(b)), with the
consequent decay of the plasma wave to $u \approx 0$, or (ii) fail
to lock the phase, resulting in $u = u_{\infty}$ after the pulse
pair. Phase locking at $\phi_0 = -\pi/2$ indicative of a transfer
to the equilibrium Branch 3 and, therefore, equilibrium
bi-stability, is never observed.

So far the effect of the plasma wave on the driver has been
neglected. Of course, the energy of the plasma wave is supplied by
the beatwave. Since the plasma wave energy changes
non-monotonically, different portions of the beatwave either lose
or gain energy. In the weakly relativistic case, the plasma energy
density $U_p \approx n_0 mc^2 u^2/2$. For concreteness, I
concentrate on the above-threshold case plotted in
Fig.~\ref{fig:beatwave_fig2} (solid line). The leading portion of
the beatwave ($-\infty < \tau < 64$) contributes energy to the
beatwave and is, therefore, depleted. If the beatwave is produced
by a laser pulse, this depletion can be described in the language
of photon deceleration, or red-shifting~\citeauthor{wilks_prl89}.
In the context of the laser beatwave the red-shifting corresponds
to the scattering of the photons from the higher frequency into
the Stokes component. Assuming equal amplitude lasers, $E_1 =
E_2$, the rate of the frequency shifting (per unit of the
propagation length) can be found as $-d\omega/dz \approx
(\omega_p^3/4 c \omega_1 a) \times d(u^2)/d\tau$. Therefore, the
laser pulse is red (blue) shifted if $d u/d\tau > 0$ ($d u/d\tau <
0$).

If the beatwave is produced by a microbunched electron beam, the
sign of $d u/d\tau$ can be related to the acceleration or
deceleration gradient of the drive electron bunch $E_z$ through
\begin{equation}\label{eq:decel}
    \frac{E_z(\tau)}{E_{\rm WB}} = \frac{\delta n_b}{n_{b0}}
    \left( \frac{1}{2 a(\tau)} \frac{d u^2}{d \tau} \right),
\end{equation}
where $E_{\rm WB}= mc\omega_p/e$ is the non-relativistic
wavebreaking electric field. Again, the sign of $d u/d\tau$
determines whether the driving bunch is accelerated or
decelerated. For a microbunched electron driver consisting of
femtosecond bunches with duration $\delta t \ll
1/\omega_p$~\cite{liu_prl98} produced by an inverse free-electron
laser $\delta n_b \sim n_{b0}$. It is estimated that in the plasma
wave decay region of the driving bunch ($64 < \tau < 112$) the
beam is decelerated at a rate of $E_z \approx 30$ GeV/m for $n_0 =
10^{19}$cm$^{-3}$. Therefore, the marriage of the microbunched
plasma wakefield accelerator and the dynamic relativistic
bi-stability concepts yields a new advanced acceleration technique
which takes advantage of the temporal drive beam structure to
produce high energy femtosecond electron beams.

Support for this work was provided by the US Department of Energy
under Contracts No.~DE-FG02-04ER54763 and DE-FG02-03ER41228.

\bibliography{beatbib}

\end{document}